\documentclass[proceedings]{JHEP} 

\usepackage{epsfig}                   

\newcommand{\lsim}{\mathrel{\mathop{\kern 0pt \rlap
  {\raise.2ex\hbox{$<$}}}
  \lower.9ex\hbox{\kern-.190em $\sim$}}}
\newcommand{\gsim}{\mathrel{\mathop{\kern 0pt \rlap
  {\raise.2ex\hbox{$>$}}}
  \lower.9ex\hbox{\kern-.190em $\sim$}}}

\conference{Trieste Meeting of the TMR Network on Physics beyond the SM}

\title{Neutrino oscillation effect on the indirect signal of
neutralino dark matter from the Earth core}

\author{Nicolao Fornengo \\
        Instituto de F\'\i sica Corpuscular -- C.S.I.C. \\
        Departamento de F\'\i sica Te\`orica, Universitat de Val\`encia \\
        c./ Dr. Moliner 50, E--46100 Burjassot, Val\`encia, Spain \\
        E-mail: \email{fornengo@flamenco.ific.uv.es}}

\abstract{We investigate the effect induced by 
          neutrino oscillation on the dark matter indirect detection 
          signal which consists in a muon neutrino flux produced by neutralino 
          annihilation in the Earth core. We consider the neutrino
          oscillation parameters relevant to the atmosferic neutrino
          deficit, both in the $\nu_\mu \rightarrow \nu_\tau$ and
	  $\nu_\mu \rightarrow \nu_s$ cases.}

\begin{document}

\section{Introduction}
Many different techniques have been proposed
for the detection of dark matter particles which could 
make up the halo of our Galaxy. Among the different
possibilities, the detection of a neutrino flux by
means of neutrino telescopes represents certainly 
an interesting tool, which is already at the level of
imposing some (mild) constraint on the particle physics
properties of the neutralino, the most interesting and
studied dark matter candidate \cite{refer}. This
particle is present in all the supersymmetric extensions
of the standard model as a linear combination
of the superpartners of the neutral gauge and higgs fields.
In the present paper we will perform our calculations
in the minimal supersymmetric extension of the standard
model (MSSM), for a definition of which we refer to Ref.\cite{ICTP}
and to the references therein quoted.

\section{Up--going muons from neutralino annihilation in the Earth}
The neutrino flux has origin from neutralino pair annihilation inside 
the Earth where these dark matter particles 
can be accumulated after having been captured by gravitational 
trapping. The differential flux is 
\begin{equation}
\Phi^0_{\stackrel{(-)}{\nu_\mu}} (E_\nu) \equiv 
\frac{dN_{\stackrel{(-)}{\nu_\mu}}}{dE_\nu} =
\frac{\Gamma_A}{4\pi d^2} \sum_{F,f}
B^{(F)}_{\chi f}\frac{dN_{f {\stackrel{(-)}{\nu_\mu}}}}{dE_\nu} \, ,  
\label{eq:fluxnu}
\end{equation}
where $\Gamma_A$ denotes the annihilation rate,
$d$ is the distance of the detector from the source (i.e. the
center of the Earth), $F$ lists the 
neutralino pair annihilation final states,
$B^{(F)}_{\chi f}$ denotes the branching ratios into
heavy quarks, $\tau$ lepton and gluons in the channel $F$.
The spectra $dN_{f {\stackrel{(-)}{\nu_\mu}} }/dE_{\nu}$ 
are the differential distributions of the (anti) neutrinos generated 
by the $\tau$ and by hadronization of quarks
and gluons and the subsequent semileptonic decays of the
produced hadrons. For details, see for instance
Refs. \cite{ICTP,noi_nuflux,altri_nuflux}. Here we only
recall that the annihilation rate depends, through
its relation with the capture rate of neutralinos in the Earth,
on some astrophysical parameters, the most relevant of which
is the local density $\rho_l$.

The neutrino flux is produced in the inner part of the Earth \cite{ICTP}
and propagates toward a detector
where it can be detected as a flux of up--going muons,
as a consequence of neutrino--muon conversion inside
the rock that surrounds the detector. A double differential
muon flux can be defined as
\begin{eqnarray}
& & \frac{d^2 N_\mu}{d E_\mu d E_\nu} = \\
& & N_A \int_0^\infty dX \int_{E_\mu}^{E_\nu} d E'_\mu 
g(E_\mu, E'_\mu; X) \; S(E_\nu, E_\mu) \nonumber \, ,
\end{eqnarray}
where $N_A$ is the Avogadro's number, 
$g(E_\mu,E'_\mu; X)$ is the survival
probability that a muon of initial energy $E'_\mu$
will have a final energy $E_\mu$ after propagating
along a distance $X$ inside the rock  and
\begin{equation}
S(E_\nu, E_\mu) = \sum_i
      \Phi_i(E_\nu) \frac{d \sigma_i(E_\nu,E'_\mu)}{d E'_\mu}
\end{equation}
where $i =  \nu_\mu, \bar\nu_\mu$
and $d \sigma_{{\stackrel{(-)}{\nu_\mu}}} (E_\nu,E'_\mu) / d E'_\mu$ is
the charged current cross--section for the
production of a muon of energy $ E'_\mu$ from 
a neutrino (antineutrino) of energy $E_\nu$.

\FIGURE[t]{
\epsfig{figure=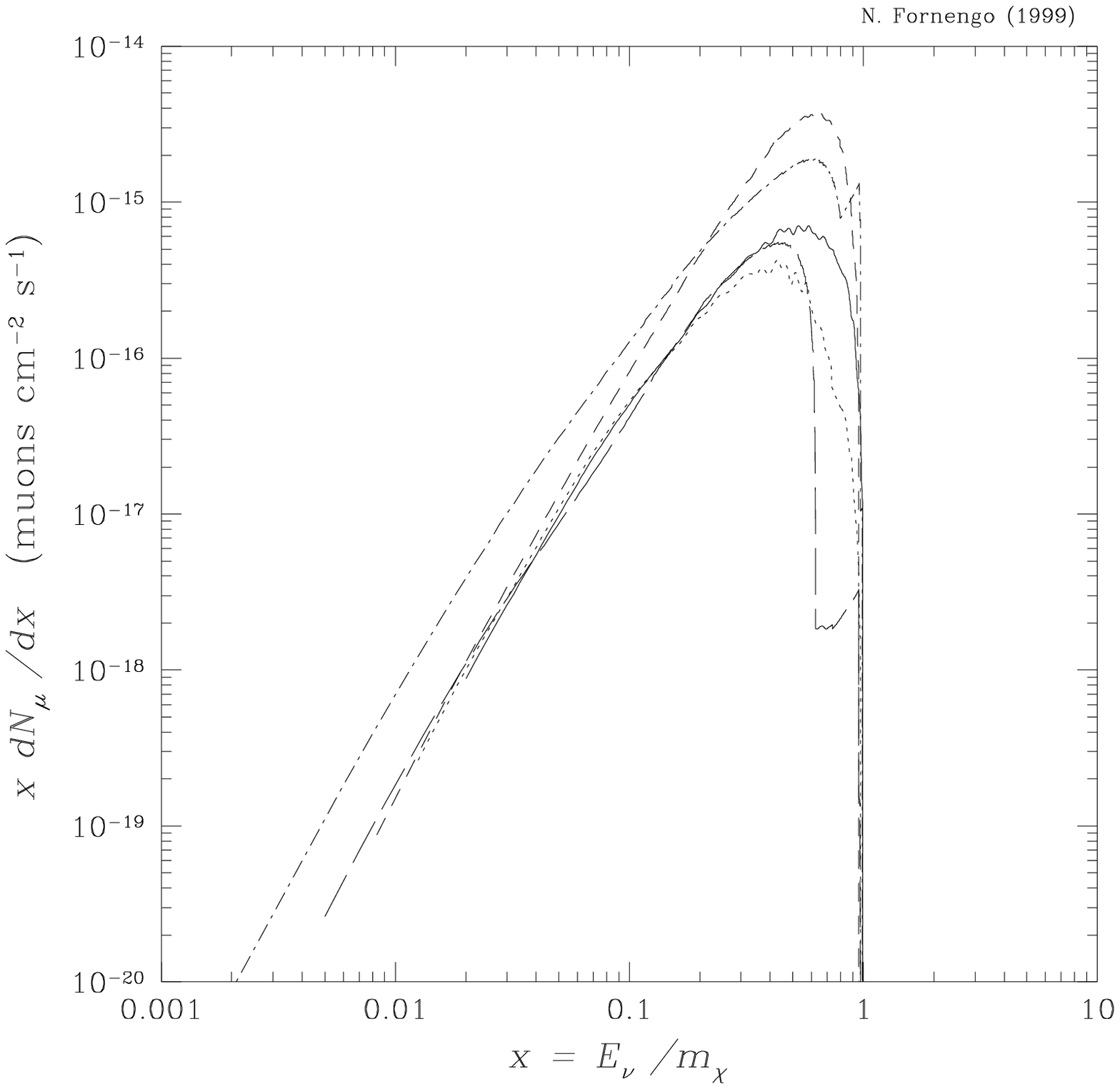,width=6.5cm,bbllx=50bp,bblly=200bp,bburx=520bp,bbury=650bp,clip=}
\caption{Muon response function $d N_\mu / d\log x$ vs.
the parent neutrino fractional energy $x = E_\nu / m_\chi$ for
neutralino annihilation in the Earth. Different curves refer to
different neutralino masses : $m_\chi = 50$ GeV (solid), 
$m_\chi = 80$ GeV (dotted), $m_\chi = 120$ GeV (shot--dashed), 
$m_\chi = 200$ GeV (long--dash), $m_\chi = 500$ GeV (dot--dashed).}
}

A useful quantity for the discussion in the following Sections is
the muon response function
\begin{equation}
\frac{d N_\mu}{d E_\nu} = \int_{E^{\mathrm{th}}}^{E_\nu}
d E_\mu \; \frac{d^2 N_\mu}{d E_\mu d E_\nu}
\end{equation}
where $E^{\mathrm{th}}$ is minimal energy for detection
of up--going muons. For SuperKamiokande and MACRO, 
$E^{\mathrm{th}} \simeq 1.5$ GeV \cite{oscill_exp}. 
The muon response
function indicates the neutrino energy range
that is mostly responsile for the up--going muon signal.
Fig. 1 shows a few examples of it, plotted as 
functions of the variable $x = E_\nu/m_\chi$, where
$m_\chi$ denotes the neutralino mass. Fig. 1 shows that
the maximum of the muon reponse happens for neutrino
energies of about $E_\nu \simeq (0.4 - 0.6) \; m_\chi$, with
a half width which extends from $E_\nu \simeq 0.1\; m_\chi$
to $E_\nu \simeq 0.8 \; m_\chi$.  

Finally, the total flux of up--going muons is defined as
\begin{equation}
\Phi_\mu = \int_{E^{\mathrm th}}^{m_\chi}
d E_\nu \; \frac{d N_\mu}{d E_\nu}
\end{equation}

\FIGURE[t]{
\epsfig{figure=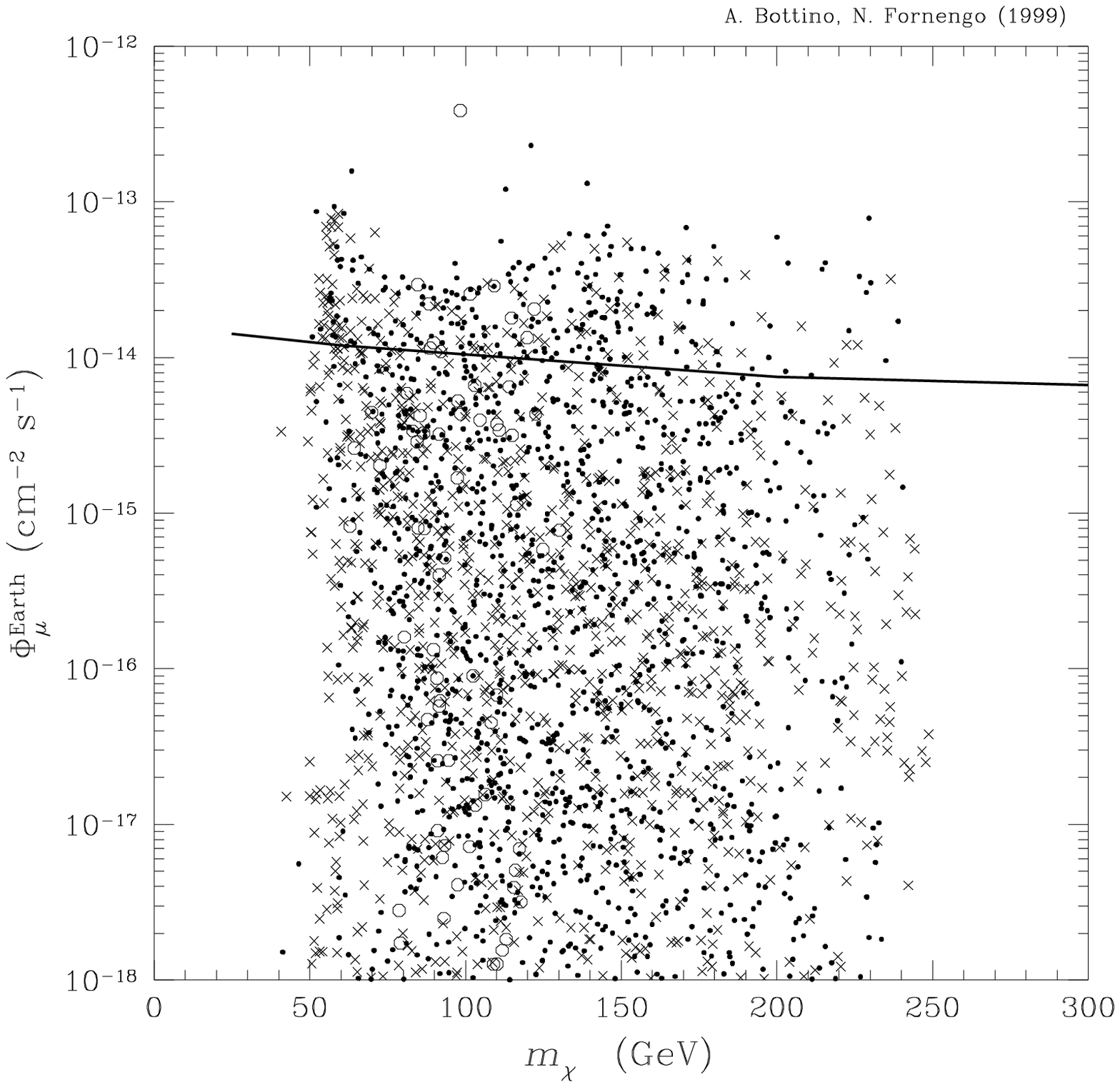,width=6.5cm,bbllx=50bp,bblly=200bp,bburx=520bp,bbury=650bp,clip=}
\caption{Flux of up--going muons $\Phi_\mu^{\mathrm{Earth}}$ from
neutralino annihilation in the Earth, plotted as a function of
$m_\chi$. The solid line denotes the present upper limit \cite{MACRO}.
Different neutralino compositions are shown with different symbols:
crosses for gauginos, open circles for higgsinos and dots for mixed
neutralinos.}
}

The natural background for these kind of
searches is represented by the flux of up--going
muons originated by the atmospheric neutrino flux.
Experimentally one searches, inside a small angular 
cone around the center of the Earth,
for a statistically significant up--going
muon excess over the muons of atmospheric $\nu_\mu$ origin.
No excess has been found so far and therefore, an upper
limit on $\Phi_\mu$ can be derived. Fig. 2 shows the
present most stringent upper limit obtained by the 
MACRO Collaboration \cite{MACRO}. In the same figure 
the theoretical calculations of $\nu_\mu$ for a scan
of the supersymmetric parameter space are also displayed.
The plot refers to  $\rho_l = 0.3$ GeV cm$^{-3}$
and is obtained by a variation of the MSSM parameters  
in the following ranges:
$20\;\mbox{GeV} \leq M_2 \leq  500\;\mbox{GeV}$,  
$20\;\mbox{GeV} \leq |\mu| \leq  500\;\mbox{GeV}$,
$80\;\mbox{GeV} \leq m_A \leq  1000\;\mbox{GeV}$,
$100\;\mbox{GeV} \leq m_0 \leq  1000\;\mbox{GeV}$,
$-3 \leq {\rm A} \leq +3,\; 1 \leq \tan \beta \leq 50$.
For further details
of the calculation, we refer to Ref. \cite{ICTP}. The comparison
of the scatter plot with the experimental upper limit would
imply that a fraction of the supersymmetric configuration
could be excluded. However, a variation of the value
of $\rho_l$ inside its range of uncertainty can lower the
theoretical prediction by about a factor of 3 \cite{ICTP}. As a consequence,
we can conservatively consider that only a small fraction of the susy 
configurations can be potentially in conflict with the experimental
upper limit, when no oscillation effect on the neutrino signal
is assumed.

\section{Neutrino oscillation effect on the up--going muon signal}

The recent data on the atmospheric neutrino deficit indicate that
the $\nu_\mu$ may oscillate, either into $\nu_\tau$ or into
a sterire neutrino $\nu_s$ \cite{oscill_exp,oscill_the}. If this
is the case, also the  $\nu_\mu$ produced by neutralino annihilations
would undergo an oscillation process. The energies involved in both
atmospheric and neutralino--produced neutrinos are the same. 
The baseline of oscillation of the two neutrino components is
different, since atmospheric neutrinos cross the entire Earth,
while neutrinos produced by  neutralino annihilation travel
from the central part of the Earth to the detector
(we recall once more that neutralinos annihilate in the core of the
Earth). On the basis of the features of the $\nu_\mu$  oscillation
which are required to fit the experimental data on atmospheric
neutrinos \cite{oscill_exp,oscill_the}, we expect that also
the neutrino flux from dark matter annihilation would be
affected. In the next Sections we will explicitely discuss
the  $\nu_\mu \rightarrow \nu_\tau$ and the
$\nu_\mu \rightarrow \nu_s$ cases, in a two neutrino mixing 
scenario \cite{Ellis}.

\subsection{$\nu_\mu \rightarrow \nu_\tau$ vacuum oscillation}
\FIGURE[t]{
\epsfig{figure=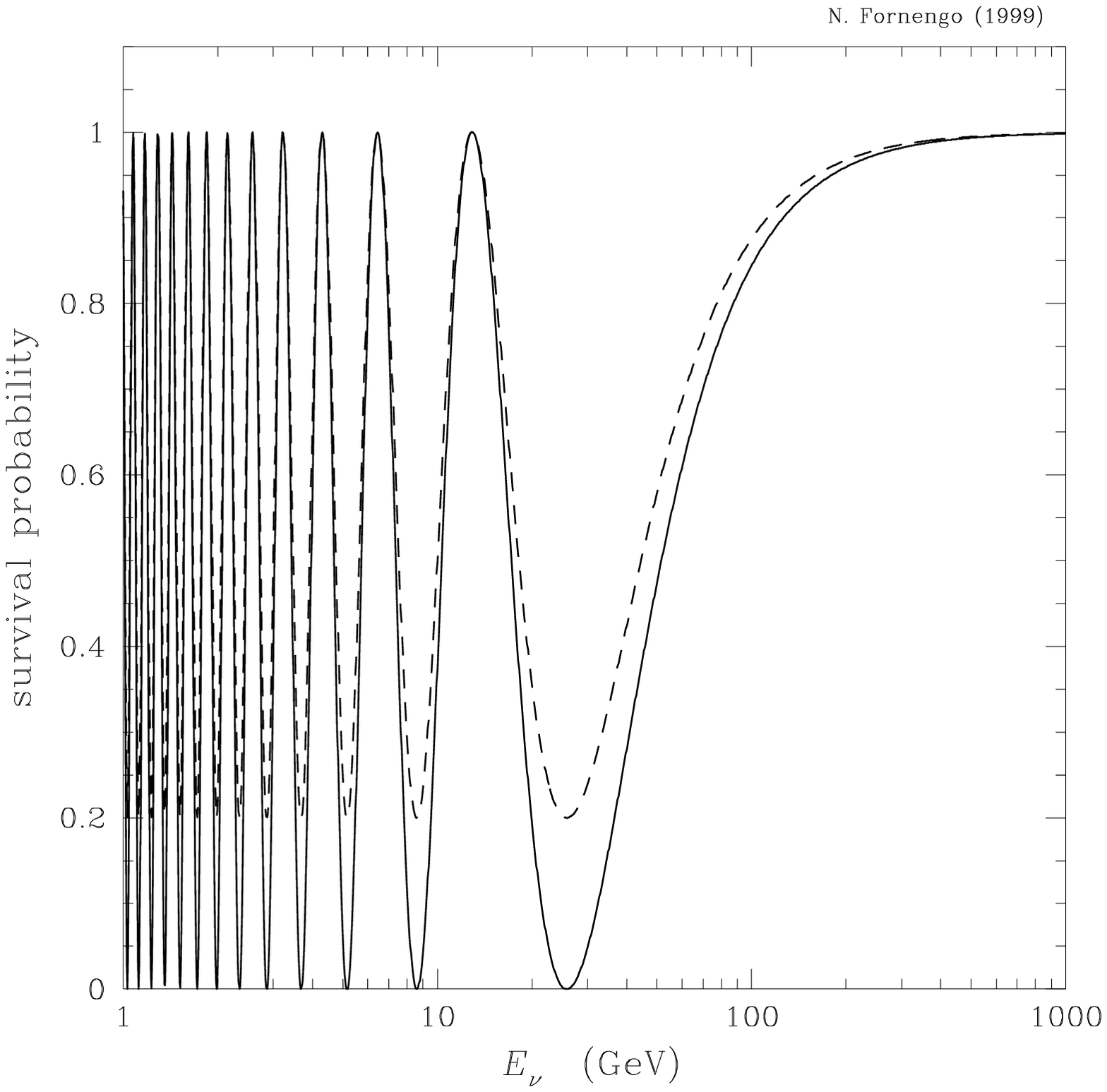,width=6.5cm,bbllx=50bp,bblly=200bp,bburx=520bp,bbury=650bp,clip=}
\caption{$\nu_\mu$ survival probability in the case of
$\nu_\mu \rightarrow \nu_\tau$ oscillation. The solid line refers to
$\sin^2 (2\theta_v) = 1$, the dashed line is for $\sin^2 (2\theta_v) = 0.8$.
In both cases, $\Delta m^2 = 5\cdot 10^{-3}$ eV $^{-2}$.}
}

\FIGURE[t]{
\epsfig{figure=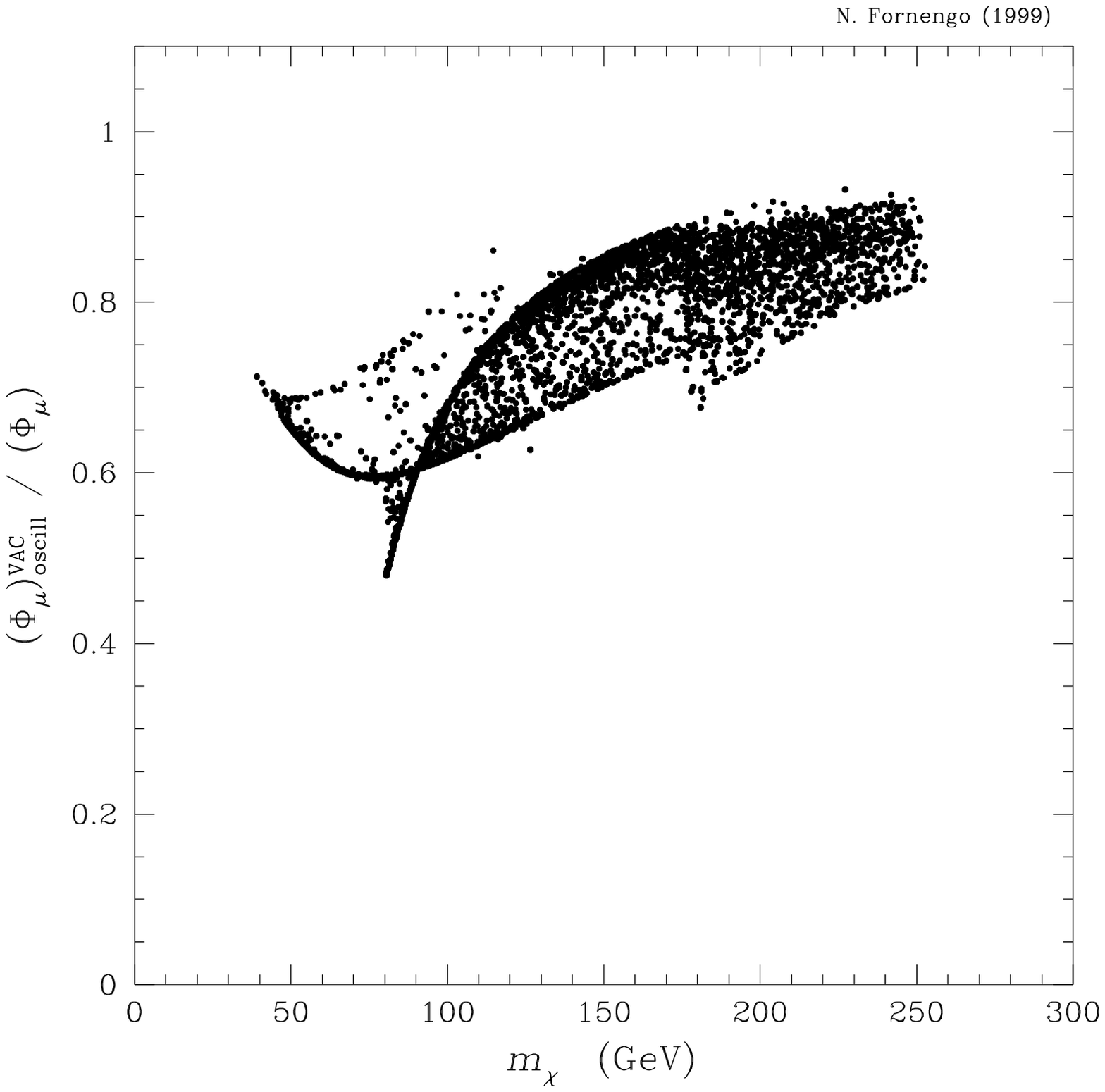,width=6.5cm,bbllx=50bp,bblly=200bp,bburx=520bp,bbury=650bp,clip=}
\caption{Scatter plot of the ratio 
$(\Phi_\mu)^{\rm VAC}_{\rm oscill}/\Phi_\mu$ vs. the neutralino
mass $m_\chi$. $(\Phi_\mu)^{\rm VAC}_{\rm oscill}$ is the up--going
muon flux in the case of $\nu_\mu \rightarrow \nu_\tau$ oscillation,
while $\Phi_\mu$ is the corresponding flux in the case of no oscillation.}
}

In the case of $\nu_\mu \rightarrow \nu_\tau$ oscillation,
the  $\nu_\mu$ flux is reduced because of oscillation, but 
we have to take into account also that neutralino annihilation 
can produce $\nu_\tau$ which in turn can oscillate into
 $\nu_\mu$ and contribute to the up--going muon flux.
The $\nu_\tau$ flux can be calculated as discussed in Sect. 1
for the $\nu_\tau$ flux, and it turns out to be always
a relatively small fraction of the  $\nu_\mu$ flux.
The muon neutrino flux at the conversion region can therefore
be expressed as
\begin{eqnarray}
\Phi_{{\stackrel{(-)}{\nu_\mu}}} (E_\nu) &=& 
\Phi^0_{{\stackrel{(-)}{\nu_\mu}}}\; 
P^{\mathrm{vac}} ({{\stackrel{(-)}{\nu_\mu}}} \rightarrow
{{\stackrel{(-)}{\nu_\mu}}}) \nonumber \\
& + & 
\Phi^0_{{\stackrel{(-)}{\nu_\tau}}}\; 
P^{\mathrm{vac}} ({{\stackrel{(-)}{\nu_\tau}}} \rightarrow
{{\stackrel{(-)}{\nu_\mu}}})
\end{eqnarray}
where the vacuum survival probability is
\begin{eqnarray}
& & P^{\mathrm{vac}} ({{\stackrel{(-)}{\nu_\mu}}} \rightarrow
{{\stackrel{(-)}{\nu_\mu}}})
\;\; = \;\; \\  
& & 1 - \sin^2(2\theta_v)\sin^2
\left (
\frac{1.27 \Delta m^2 (\mathrm{eV}^2) R(\mathrm{Km})}
{E_\nu (\mathrm{GeV})} \nonumber
\right )
\label{vac}
\end{eqnarray}
where $\Delta m^2$ is the mass square difference of the
two neutrino mass eigenstates, $\theta_v$ is the mixing angle
in vacuum and $R$ is the Earth's radius. 
Fig. 3 shows the survival probability for two different values of
the neutrino oscillation parameters. Smaller (larger) values of 
$\Delta m^2$ have the effect of shifting the curves to the left (right).
Comparing Fig. 1 with Fig. 3, we notice that the reduction of the
up--going muon flux is stronger when there is matching between the
the energy 
$E_\nu^1 \simeq 5.2 \cdot 10^{-3} \Delta m^2 (\mathrm{eV}^2)$ 
of the first (from the right) minimum of the 
survival probability and the energy 
$E_\nu \simeq 0.5 m_\chi$ 
which is responsible for most of the muon response in the detector.
This implies that a maximum reduction of the signal could
occur for neutralino masses of the order of 
$m_\chi (\mathrm{GeV}) \simeq 10^4 \Delta m^2 (\mathrm{eV}^2)$.
The $\nu_\tau \rightarrow \nu_\mu$ oscillation makes the
reduction of the muon flux less severe, but it is not able
to completely balance the reduction effect because the
original $\nu_\tau$ flux at the source is sizeably smaller than
the $\nu_\tau$ flux. Therefore, the overall effect of
the neutrino oscillation is to reduce the up--going muon
signal. This effect is summarized in Fig. 4, where the
ratio between the up--going muon signals in the presence
and in the absence of oscillation are plotted as a function
of the neutralino mass. The susy parameter space has been
varied in the same ranges quoted for Fig. 2. We notice
that the strongest effect is present for light neutralinos,
since in this case the muon flux is mostly produced from
neutrinos whose energy is in the range of maximal suppression
for the oscillation phenomenon. The effect is between
0.5 and 0.8 for $m_\chi \lsim 100$ GeV.
On the contrary, the fluxes for larger masses are less 
affected, and the reduction is less than about 
20\% for  $m_\chi \gsim 200$ GeV.

\subsection{$\nu_\mu \rightarrow \nu_s$ matter oscillation}
\FIGURE[t]{
\epsfig{figure=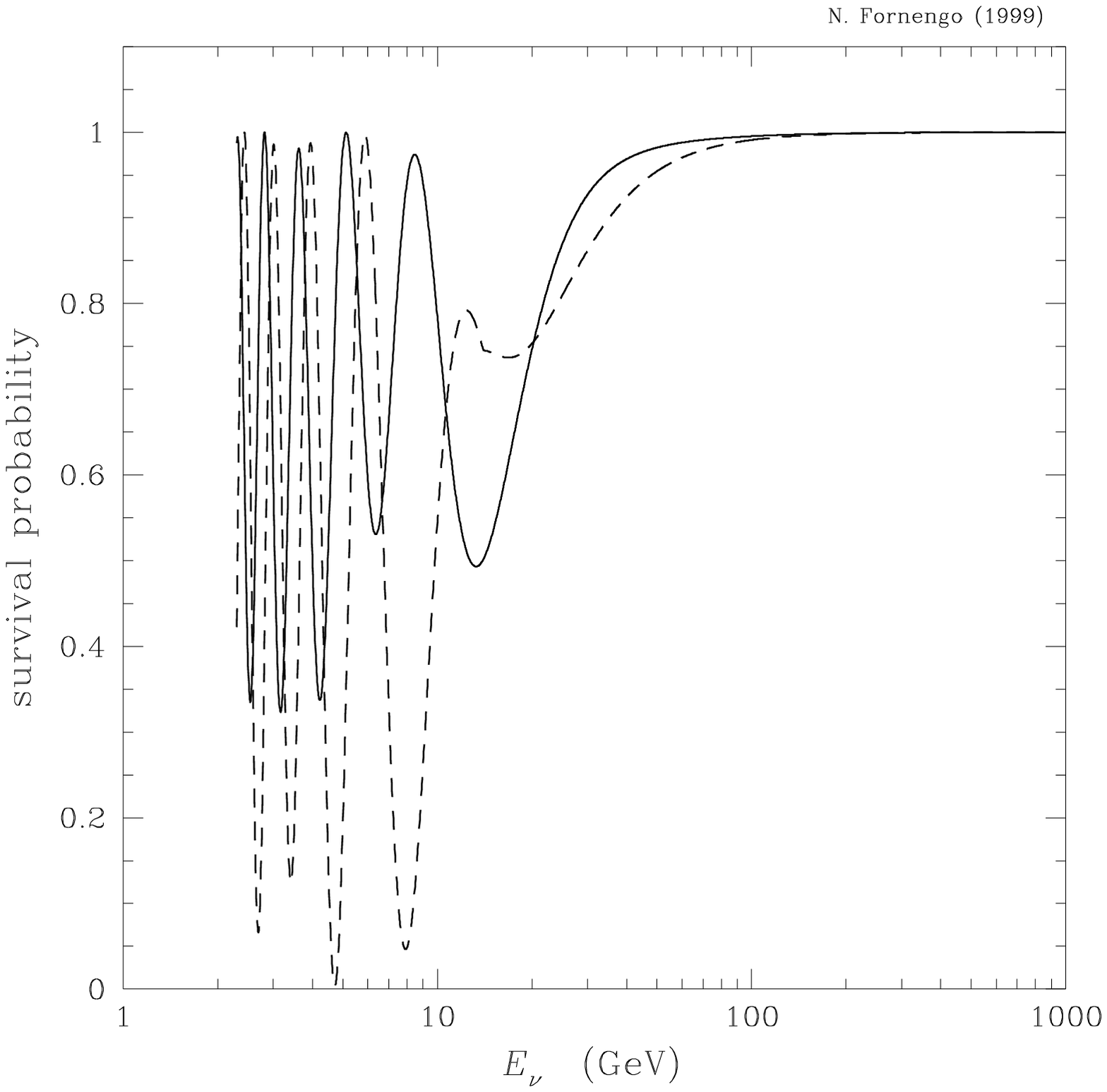,width=6.5cm,bbllx=50bp,bblly=200bp,bburx=520bp,bbury=650bp,clip=}
\caption{$\nu_\mu$ survival probability in the case of
$\nu_\mu \rightarrow \nu_s$ oscillation, for $\sin^2 (2\theta_v) = 0.8$
and $\Delta m^2 = 5\cdot 10^{-3}$ eV $^{-2}$. The solid line refers to
neutrinos, the dashed line is for antineutrinos.}
}

\FIGURE[t]{
\epsfig{figure=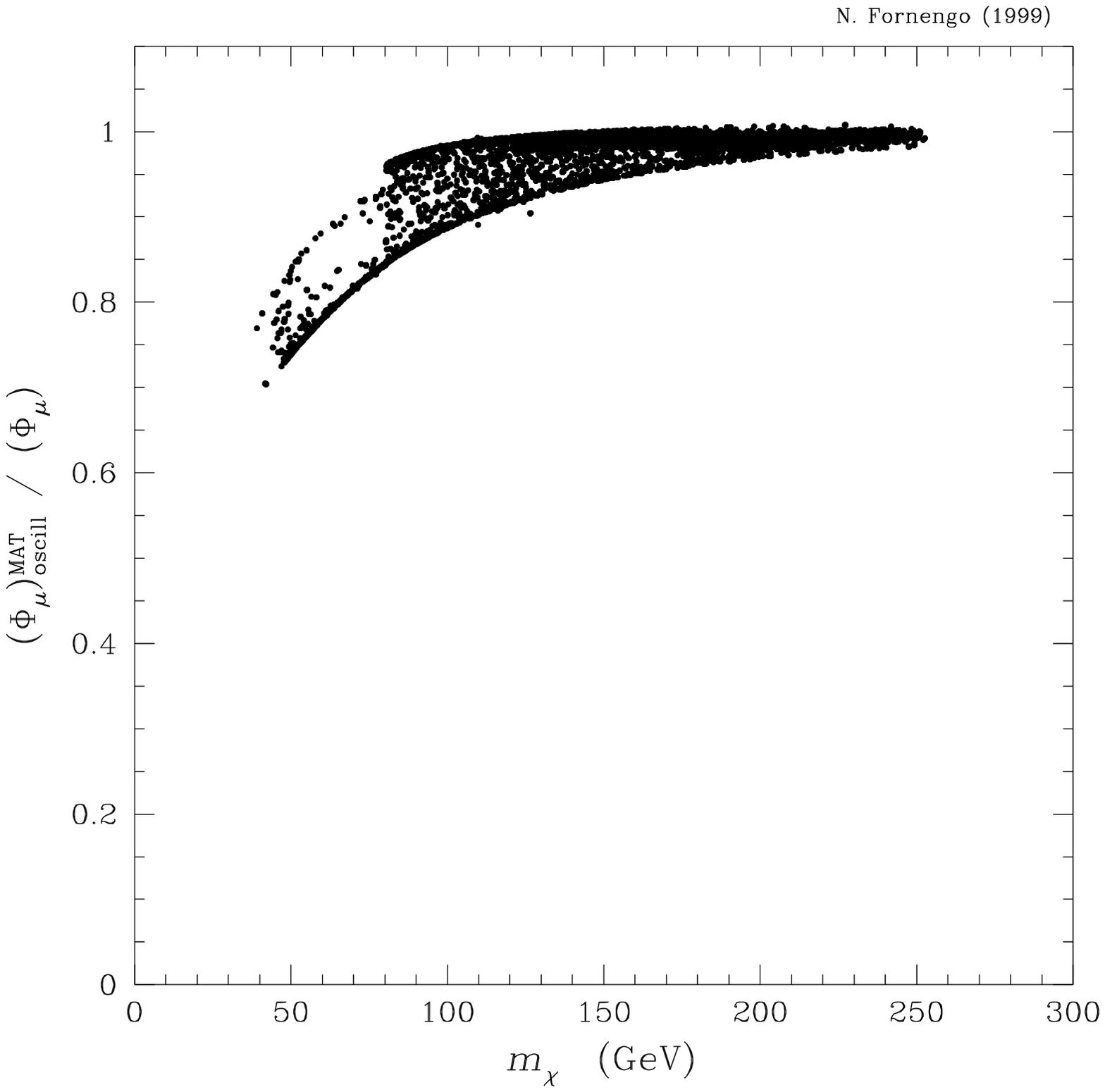,width=6.5cm,bbllx=50bp,bblly=200bp,bburx=520bp,bbury=650bp,clip=}
\caption{Scatter plot of the ratio 
$(\Phi_\mu)^{\rm MAT}_{\rm oscill}/\Phi_\mu$ vs. the neutralino
mass $m_\chi$. $(\Phi_\mu)^{\rm MAT}_{\rm oscill}$ is the up--going
muon flux in the case of $\nu_\mu \rightarrow \nu_s$ oscillation,
while $\Phi_\mu$ is the corresponding flux in the case of no oscillation.}
}

In the case of $\nu_\mu \rightarrow \nu_s$ oscillation, 
the neutrino flux is simply
\begin{equation}
\Phi_{{\stackrel{(-)}{\nu_\mu}}} (E_\nu) = 
\Phi^0_{{\stackrel{(-)}{\nu_\mu}}}\; 
P^{\mathrm{mat}} ({{\stackrel{(-)}{\nu_\mu}}} \rightarrow
{{\stackrel{(-)}{\nu_\mu}}})
\end{equation}
and no $\nu_\mu$ regeneration is possible from the
sterile neutrino. In this case, the effective potential
of $\nu_\mu$ and $\nu_s$ inside the Earth are different
and we have to solve the evolution equation for propagation
in the core and in the mantle. Neutrinos
(produced in the center of the Earth) cross once half
of the core and once the mantle. By considering both
core and mantle as of constant density, we can express the
survival probability as \cite{akhmedov, Kim}
\begin{eqnarray}
& & P^{\mathrm{mat}} ({{\stackrel{(-)}{\nu_\mu}}} \rightarrow
{{\stackrel{(-)}{\nu_\mu}}}) \,\,=\,\,  \\
& & \left [
U(\theta_c) D(\phi_c) U^\dagger (\theta_c - \theta_m)
 D(\phi_m) U^\dagger (\theta_m)
\right ]_{\mu\mu} \nonumber
\label{mat}
\end{eqnarray}
where $U$ is the $2 \times 2$
neutrino mixing matrix, $\theta_a$ 
($a=c,m$ for core and mantle, respectively) are 
the effective mixing angles in matter and they are related
to the vacuum mixing angle $\theta_v$ as
\begin{equation}
\sin^2 (2\theta_a) = \frac{\sin^2 (2\theta_v) \xi_a^2}
{\left [
       {(\xi_a \cos(2\theta_v) + 1)^{2} +
\xi_a^2 \sin^2 (2\theta_v)}
\right ]}
\end{equation}
with $\xi_a = \Delta m^2 / (2E_\nu V_a)$;
$V_a = \pm G_F N_n^a / \sqrt{2}$ is the matter 
potential in a medium of number density $N_n^a$
for neutrinos ($+$) and antineutrinos ($-$).
In Eq.(3.4), $D$ is the evolution
matrix $D_{ij}(\phi_a) = \delta_{ij} d_j^a$,
where $d_1^a = 1$, $d_2^a = \exp(i \phi_a)$ and
\begin{equation}
\phi_a = V_a R_a 
\left [
(\xi_a \cos(2\theta_v) + 1)^2 +
\xi_a^2 \sin^2 (2\theta_v)
\right ]^{1/2}
\end{equation}

In Fig. 5 an example of the $\nu_\mu$ and $\bar \nu_\mu$  
survival probability is given for representative
values in the range allowed by the fits on the
atmospheric neutrino data \cite{oscill_exp,oscill_the}:
$\Delta m^2 = 5\cdot 10^{-3}$ eV$^2$
and  $\sin^2 (2\theta_v) = 0.8$. {}From Fig. 5
and the previous discussion relative to Fig. 3, we 
expect that in the case of  $\nu_\mu \rightarrow \nu_s$
the reduction of the muon signal is significantly
less severe than in the case of   $\nu_\mu \rightarrow \nu_\tau$.
In fact, in these case the minima of the survival probability
occur for lower neutrino energies, and threfore the
oscillation can affect only muon fluxes originated by very
light neutralinos. This is manifest in Fig. 6, where
the ratio of the up--going muon fluxes in presence and
absence of oscillation are shown. In this case, the reduction 
of the signal is always less than 30\%. This maximal reduction
occurs for neutralino masses lower that about 80 GeV. For
larger masses, the up--going muon flux is almost unaffected.

\section{Conclusions}
We have discussed the effect on the up--going muon
signal from neutralino annihilation in the Earth, in the
case that the $\nu_\mu$ flux produced by neutralinos would oscillate
as indicated by the data on the atmospherice neutrino deficit.
While the experimental upper limit is, at present, practically
not affected by the possibility of neutrino 
oscillation \cite{MACRO}, the
theoretical predictions are reduced in the presence of
oscillation. With the oscillation parameters deduced
{}from the fits on the atmospheric neutrino data,
the effect is always larger for lighter neutralinos.
In the case of $\nu_\mu \rightarrow \nu_\tau$
the reduction is between 0.5 and 0.8 for 
$m_\chi \lsim 100$ GeV and less than about
20\% for  $m_\chi \gsim 200$ GeV.
In the case of  $\nu_\mu \rightarrow \nu_s$,
the reduction of the signal is up to 30\% for neutralino 
masses lower that about 80 GeV and smaller than 10\% for
heavier neutralinos.

\acknowledgments
I wish to thank Sandro Bottino for very stimulating and
interesting discussions about the topic of this paper.
This work was supported by DGICYT under grant number 
PB95--1077 and by the TMR network grant ERBFMRXCT960090 of 
the European Union.

\end{document}